# Angle dependent field-driven reorientation transitions in uniaxial antiferromagnet MnBi$_2$Te$_4$ single crystal

Ning Cao,[1,&] Xue Chen,[2,&] Xinrun Mi,[2] Saisai Qiao,[1] Liyu Zhang,[2] Kunling Peng,[1] Mingquan He,[2] Aifeng Wang,[2] Yisheng Chai,[2*] Xiaoyuan Zhou[1*]

[1]*Center for Quantum Materials & Devices and College of Physics, Chongqing University, Chongqing 401331, China*

[2]*Low Temperature Physics Laboratory, College of Physics, Chongqing University, Chongqing 401331, China*

[*]Corresponding authors: yschai@cqu.edu.cn, xiaoyuan2013@cqu.edu.cn

[&]These authors contribute equally

## Abstract

MnBi$_2$Te$_4$, a two-dimensional magnetic topological insulator with a uniaxial antiferromagnetic structure, is an ideal platform to realize quantum anomalous Hall effect. However, the strength of magnetic interactions is not clear yet. We performed systematic studies on the magnetization and angle dependent magnetotransport of MnBi$_2$Te$_4$ single crystal. The results show that the direction of the magnetic field has significant effects on the critical field values and magnetic structure of this compound, which leads to different magnetotransport behaviors. The field-driven reorientation transitions can be utilized to estimate the AFM interlayer exchange interaction coupling and uniaxial magnetic anisotropy *D*. The obtained Hamiltonian can well explain the experimental data by Monte Carlo simulations. Our comprehensive studies on the field-driven magnetic transitions phenomenon in MnBi$_2$Te$_4$ provide a general approach for other topological systems with antiferromagnetism.

# I. INTRODUCTION

In recent decades, the combination of topology and condensed matter physics has become a new research field. Through theoretical predictions and experiments, a variety of topological materials have been discovered, such as topological insulator (TI) [1, 2], Dirac semimetal [3, 4], and Weyl semimetal [5, 6] etc. A series of related emerging phenomena in physics, such as quantum anomalous Hall effect (QAHE) and Majorana fermions was predicted. QAHE was first realized in Cr- or V- doped $(Bi, Sb)_2Te_3$ magnetic topological thin films exhibiting strong inhomogeneity [7, 8]. Recently, an antiferromagnetic (AFM) TI $MnBi_2Te_4$, being cleaved down to several layers, a nearly saturated magnetic state is realized and zero field QAHE up to $0.97h/e^2$ is observed [9]. The demonstration of the fully quantized QAHE still requires high enough magnetic field to enter ferromagnetic (FM) state. When Mn-Te is intercalated with multiple $Bi_2Te_3$, a series of interesting structures $MnTe(Bi_2Te_3)_n$ will be formed [10-12]. Applying a magnetic field with in-plane changes the magnetic structure of $MnTe(Bi_2Te_3)_n$ ($n = 1, 2, 3$) and leads to the first material platform for a higher-order Möbius insulator with a Möbius twist in its topological surface state [13].Therefore, it is crucial to understand the magnetic phase transitions of this compound to realize exotic topological states in this system.

For $n = 1$, $MnBi_2Te_4$ is a rhombohedral structure with a space group of $R$-$3m$ [10, 14] and is a Van der Waals layered material composed of TI $Bi_2Te_3$ and MnTe intercalated, which is stacked in a sequence of septuple layers (SLs) Te-Bi-Te-Mn-Te-Bi-Te in the $z$-axis, as shown in Fig. 1(a). The gap of $MnBi_2Te_4$ is about 70-88 meV [15, 16], much larger than that of Cr- or V- doped $(Bi, Sb)_2Te_3$ TI thin films. The magnetic moments of Mn atoms in a SL are in parallel with each other and form a FM order with the out-of-plane easy axis, while the interlayer exchange coupling between neighboring SLs is AFM, generating a thre e-dimensional A-type AFM order [16-18]. To realize a zero-field ideal QAHE in a few layers, a FM state should be stabilized without magnetic field in this system. A series of experimental studies on magnetic interactions and phase transitions on the bulk samples have been performed, with the help of various model Hamiltonians and theoretical approaches. The obtained exchange interactions are not fully consistent among different experimental techniques and models. M. M. Otrokov *et al.* has used first principle calculation to obtain the single-site spin stiffness coefficient $J_0^\perp$ (interlayer coupling constant) and the magnetic anisotropy energy which are equal to -0.022 meV/$\mu_B^2$ and 0.225 meV per Mn atom, respectively [16]. R. J. McQueeney *et al.* provided an *SD* (*S* = 5/2, *D* is uniaxial anisotropy parameter) and -$SJ_c$ value of 0.07-0.1 meV and 0.05-0.09 meV, respectively, using the Heisenberg model. A rough estimation of *SD* = 0.12 meV and -$SJ_c$ = 0.055 meV is obtained from inelastic neutron scattering data by considering the spin gap and the bandwidth of interlayer excitation [19]. We noticed that, from the characteristic magnetic field and angular dependent magnetization behaviors of an AFM, the magnetic states and material parameters can be estimated, as demonstrated in layered antiferromagnet orthorhombic ($C_2H_5NH_3$)$_2$CuCl$_4$ [20]. Therefore, similar approach can be applied in this system by measuring angle dependent properties and characteristic field values to quickly identify the dominating magnetic interactions.

In this paper, we successfully grew high quality MnBi$_2$Te$_4$ single crystal. Apart from conventional magnetization data, the angle of azimuthal angle ($\varphi$) and polar angle ($\theta$) dependent magnetoresistance measurements reveal a very weak in-plane hexagonal anisotropy and strongly field-driven reorientation transitions, respectively. From the above measurements, a $J_c$-*D* model Hamiltonian with AFM interaction $J_c$ = 0.105 meV and uniaxial anisotropy *D* = 0.033 meV is established. Monte Carlo simulation on this model can well reproduce the polar phase diagram drawn from all the magnetic and electrical measurements. These findings would advance understanding of the uniaxial AFM in MnBi$_2$Te$_4$.

## II. MATERIALS AND METHODS

The precursors Bi$_2$Te$_3$ and MnTe were synthesized by a solid-state reaction. High purity materials (Mn slices, 99.99%; Bi lumps, 99.999%; Te chunks, 99.999%) were mixed together according to the ratio Mn:Te = 1:1, Bi:Te = 2:3, and were put into alumina crucible (prevent material reaction with quartz tube) and then sealed in a quartz tube. The sealed quartz tubes were heated to 1100 °C (MnTe) and 900 °C ($Bi_2Te_3$) in 10 h, kept at that temperature for 24 h, and subsequently cooled to room temperature. The obtained polycrystalline MnTe and $Bi_2Te_3$ were mixed together according to the ratio MnTe:$Bi_2Te_3$ = 3:17 and loaded into a 3 ml alumina growth crucible and then sealed in a quartz tube. The quartz tube was slowly heated to 650 °C in 10 h and held for 12 h, and then slowly cooled to 595 °C at a rate of 1 °C/h. The extra flux $Bi_2Te_3$ was removed by the centrifugation process. The obtained luster crystals with size of a few millimeters were checked using X-ray diffractometer (PANalytical) with Cu $K_\alpha$ ($\lambda$ = 1.54056 Å) radiation. The magnetotransport of $MnBi_2Te_4$ was obtained on Physical Properties Measurement System (PPMS, Quantum Design Dynacool) and the resistivity $\rho_{xx}$ and $\rho_{zz}$ were measured by a standard four-probe method. For the out-of-plane resistance measurement of $\rho_{zz}$, a thick single crystal was cut to produce a needlelike sample with the long side along the $z$-axis with about 5° uncertainty in order to minimize the contribution of the in-plane resistivity component. The magnetism was measured by the VSM option on PPMS. Monte Carlo simulations are performed on bulk $MnBi_2Te_4$ with a 10 layers 1D spins with periodic boundary conditions. Monte Carlo simulations are run with the field pointed in the $yz$ plane using 30000 Monte Carlo steps per field/angle to equilibrate the structure and an additional 30000 steps to calculate the average magnetic properties. To be consistent with experimental data, we begin the simulations with $T$ = 0.1 K and ramp the field up or rotate the angle in equal steps, using the final state from the previous field as the initial state for the current field.

## III. RESULTS AND DISCUSSIONS

As shown in Fig. 1(b), the obtained high quality single crystal shows sharp (00$l$), $l$ = 3$n$, X-ray diffraction peaks, indicating that there is no residual flux $Bi_2Te_3$ on the surface. From the corresponding $R$-3$m$ space group, a lattice constant of $c \approx$ 40.730 Å can be calculated and consistent with previous reports of 40.919 Å [10] for $MnBi_2Te_4$. Both in-plane and out-of-plane magnetization ($M$) for $MnBi_2Te_4$ single crystal as a function of temperature ($T$) are measured under dc magnetic field ($\mu_0 H$ = 0.2 T). The temperature dependence of the magnetic susceptibility $\chi = M/H$ in field cooling process is summarized in Fig. 1(c). When the temperature is above $T_N$ (≈ 25 K), the inverse magnetic susceptibility shows a linear dependence of temperature, which follows the Curie–Weiss (CW) law $\chi = C/(T+\theta_{CW})$. We fit the inverse magnetic susceptibility above 150 K, the effective moment per formula is 5.75 $\mu_B$ and $\theta_{CW}$ = -24.6 K for $H//z$, close to the theoretical value of Mn$^{2+}$ ions with a 3d$^5$, $2\sqrt{S(S+1)}\mu_B$=5.92$\mu_B$ with $S$ = 5/2. Negative Weiss temperature indicates predominately AFM interaction among Mn$^{2+}$ spins, and the deviation from the CW law near $T_N$ possibly arises from strong spin fluctuations around magnetic phase transition [21]. In addition, the effective moment per formula is 5.17 $\mu_B$ and $\theta_{CW}$ = 3.8 K for $H//xy$. The discrepancy between CW behaviors of in-plane and out-of-plane directions may come from magnetic anisotropy [21]. Below $T_N$, the out-of-plane $\chi$ decreases with decreasing $T$ while that of the in-plane one is almost a constant, verifying the AFM configuration along $z$-axis.

Figure 1(d) present the temperature dependence of the in-plane resistivity $\rho_{xx}$ ($T$) from 2 K to 200 K. It shows a metallic behavior that decreases as the temperature decreases, and a distinct peak around the $T_N$. Below $T_N$, the resistivity drastically drops with temperature. The sharp peak feature possibly caused by enhanced electron scattering from spin fluctuations and the sharp decrease below $T_N$ reflects the weakened spin-disorder scattering due to an onset of long-range AFM magnetic order [22]. It is corroborated that a 9 T magnetic field along $z$-axis can destroy AFM order and totally suppress the peak and magnitude in $\rho_{xx}$ below 100 K, as shown in Fig. 1(d) and consistent with previous report [15].

Figures 2(a) and 2(b) show the magnetization curves measured for MnBi$_2$Te$_4$ at different temperatures with $H//z$ and $H//xy$, respectively. For $H//z$, the kink behavior appears near 3.3 T ($H_{c1}$) at 2 K and disappears above 25 K, indicating that a spin-flop (SF) transition, and shows saturation behavior at about 7.6 T ($H_s$) when the SF state is fully polarized to the FM state. In contrast, the $M$-$H$ curves of $H//xy$ increase linearly with the magnetic field and no obvious saturation behaviors can be observed up to 9 T below 20 K, indicating that the AFM state is gradually polarized to the FM state with $H$ increasing. We extrapolate the 2 K data up to the saturation value of $M \approx 4.2$ μ$_B$/f.u. and obtained a μ$_0H \approx 10.5$ T ($H_s$). At 20 K, a μ$_0H \approx 6.8$ T can be found without showing any metamagnetic transition. The magnetic phase diagrams of MnBi$_2$Te$_4$ for $H//z$ and $H//xy$ can be drawn from the above curves and the corresponding arrows represent the magnetic order formed by the Mn FM layers, showing AFM, SF, and FM orders, respectively.

To describe the external fields dependent spin configurations, we apply the following Hamiltonian by considering a local-moment Heisenberg model by including the AFM interlayer exchange interaction coupling $J_c<0$, uniaxial magnetic anisotropy $D$ and Zeeman energy [20, 23]:

$$H = -J_c \sum_{ij} \boldsymbol{S}_i \cdot \boldsymbol{S}_j - D \sum_i S_{i,z}^2 - g\mu_B \boldsymbol{H} \cdot \sum_i \boldsymbol{S}_i \qquad (1)$$

Where, for the hexagonal lattice, we omit the fourth and sixth-order terms since the second-order term with constant $D$ plays the dominant role for the spin orientations. We could then use the above magnetization data of $H//z$ and $H//xy$ to determine all model parameters. At 2 K, the SF and saturation fields μ$_0H_{c1} = 3.3$ T and μ$_0H_s = 7.6$ T with $H//z$ and μ$_0H_s = 10.5$ T in $H//xy$. Within the above model, these critical fields are given by the expressions: $g\mu_B H_{c1} = 2S\sqrt{(2|J_c|-D)D}$, $g\mu_B H_s(\theta=0°) = 2S(2|J_c|-D)$ and $g\mu_B H_s(\theta=90°) = 2S(2|J_c|+D)$ (where $g \approx 2$) [19]. The best fit to the experimental data can provide parameters of $J_c = 0.105$ meV and $D = 0.033$ meV. Moreover, by using Monte Carlo simulations with fitted $J_c$ and $D$ value, the $M-H$ curves along $H//z$ and $H//xy$ at 2 K can be well produced with critical field values of μ$_0H_{c1} = 3.4$ T, μ$_0H_s = 7.6$ T and μ$_0H_s = 10.5$ T. In addition, the $H//z$ data shows clear hysteresis behavior at $H_{c1}$ [24]. However, no hysteresis behavior can be found in the experiments, probably due to nucleation processes. Therefore, we use the average value between $H$ increase and decrease scans, which is 3.4 T of $H_{c1}$ to compare with that in experiments.

To probe the in-plane magnetic anisotropy of MnBi$_2$Te$_4$, we prepared a resistivity sample with interlayer configuration by measuring the $\rho_{zz}$. The in-plane magnetic field dependent resistivity $\rho_{zz}$ ($H$) along selected $\varphi$ at 2 K, as shown in Figs. 3(a) and (c). $\rho_{zz}$ ($H$) steadily decreases up to 9 T without any negligible phase transition. For $H//xy$, no $H_{c1}$ exists and $H_s$ is higher than 9 T at 2 K so that there is no kink in the $\rho_{zz}$. The resistivity curve is almost identical for every $\varphi$ value. The isotropic magnetoresistance in $\varphi$ dependent $\rho_{zz}$ justifies our assumption of negligible in-plane hexagonal anisotropy in the Hamiltonian to determine the spin configurations.

As the $H//xy$ and $H//z$ leads to different $H$-$T$ phase diagrams, it is meaningful to find out how the critical fields evolve when magnetic field continuously changes from in-plane to out-of-plane directions, as schematically shown in Fig. 3(b). We then performed the magnetoresistance measurement to characterize the transition fields. Firstly, $H$ dependence of the $\rho_{zz}$ at 2 K at selected $\theta$ in the range of 0° to 90° are measured, as shown in Fig. 3(d). In contrast with the smooth decrease of $\rho_{zz}$ in $H//xy$ ($\theta$ = 90°) configuration, $\rho_{zz}$ shows a drastic jump at 3.4 T and a kink at 7.6 T for $H//z$ ($\theta$ = 0°), corresponding to $H_{c1}$ and $H_s$ values in Fig. 2(a) respectively. In other in-between angles, $\rho_{zz}$ near $H_{c1}$ becomes broader with increasing $\theta$ while the kink field at $H_s$ gradually disappears above 45° and increases with increasing $\theta$. In the field range $H<H_{c1}$, $\rho_{zz}$ is nearly always smaller in larger $\theta$ values, indicating that the in-plane magnetic field component can strongly tune the AFM configuration than that of the out-of-plane one. In the field range $H_{c1}<H<H_s$, the magnetic order is SF that $\rho_{zz}$ ($H$) exhibit concave behavior and is most obvious when the tilting angle $\theta$ is 0°. In the field range $\mu_0H>7$ T, the $\rho_{zz}$ is weakly anisotropic with slightly larger along $\theta$ = 0° direction.

Secondly, to directly reveal such anisotropic in polar rotation, $\theta$ dependence of the $\rho_{zz}$ at 2 K at selected $H$ in the range of 1 to 9 T are measured, as shown in Fig. 3(e). In the field range $H<H_{c1}$ of 1 T and 3 T, as the magnetic field rotates from out-of-plane to in-plane, the resistivity decreases gradually. In the 5 T, which is slightly above $H_{c1}$, $\rho_{zz}$ ($\theta$) gradually increases as $H$ rotated from out-of-plane to in-plane. When the magnetic field increases to 7 T, $\rho_{zz}$ is almost isotropic with very weak kink around $\theta$ = 0°, which is also revealed in $H$ sweep data in Fig. 3(d) that they almost come across the same value around 7 T. Under 9 T, the resistivity data appear several kinks (-54°, 54°, 126°, 234°), indicating that the $H_s$ field is just reached along those directions. Between -54°

- 54° including *z*-axis, the spin configuration must be FM while between 54° - 126° including *xy* plane, the magnetization should not be saturated. The detailed angle and field dependent spin configurations will be calculated in the following based on the Heisenberg model in Eq. (1).

Figure 3(f) shows the $\theta$ dependence of the $\rho_{zz}$ under 9 T at selected temperature. The 2 K data is the same as that in Fig. 3(e). At 10 K, $\rho_{zz}(\theta)$ changes almost smoothly by reaching the maximum and minimum at *z*-axis and close to *xy* plane, respectively, as the magnetic order is FM for *H* along *z*-axis but is not along *xy* plane (Figs. 2(c) and 2(d)). At 20 K, $\rho_{zz}(\theta)$ changes smoothly with clear 2-fold anisotropy due to the FM state along all the directions. Above 20 K, the 2-fold anisotropy is gradually weakened with increasing temperature up to 50 K where there is no long-range magnetic order.

We now turn to the theory to determination of the critical field values when magnetic field continuously changes from in-plane to out-of-plane directions. The analytical expression for critical field $H_{c1}$ and saturation field $H_s$ for tilting *H* is

$$g\mu_B H_{c1} = 2S\sqrt{D[2|J_c| + D - 2D\cos^2(\theta)]}\cos(\theta) \qquad (2)$$
$$g\mu_B H_s = 2S[2|J_c| - D\cos(2\theta)]$$

if we assume that the *M* is parallel to *H* around the critical fields. This is not always true due to the existence of magnetic anisotropy. To deduce the accurate values of $H_{c1}$ and $H_s$ at each angle $\theta$, numerical Monte Carlo simulations based on the above $J_c$-*D* model are applied. To simplify further the calculation, we restrict the spatial orientation of the magnetization vectors within *yz*-plane [20]. Such assumption always holds true when *H* spans the *yz*-plane for uniaxial AFM in hexagonal crystal with very weak *xy*-plane anisotropy.

As shown in Fig. 4, we calculated the *H* dependence of the two representative spins $S_1$ and $S_2$ with positive and negative $S_z$ at $\mu_0 H = 0$ T, respectively at selected $\theta$ in the range of 0° to 90°. At $\theta = 0°$ (*H*//*z*), $S_z$ and $S_y$ show drastic jumps around 3.4 T with clear hysteresis behavior, corresponding to $H_{c1}$ value in Fig. 2(a). With increasing $\theta$ from 5° up to 30°, $S_z$ and $S_y$ around $H_{c1}$ becomes broader without hysteresis. The obtained average $H_{c1}$ values decreases with increasing $\theta$. At 45°, $H_{c1}$ cannot be reliably defined while $H_s$ = 8.9 T, which is almost consistent with the field value of 8.4 T from the magnetoresistance data in Fig. 3(d). The $S_1$ will rotate in counterclockwise first, then continuously rotate in clockwise until parallel to the $H$ while $S_2$ rotate continuously passing positive $y$-axis until parallel to the $H$. Above 70°, with increasing $H$, no backward rotation of $S_1$ is observed. In the whole angular range, the saturation field $H_s$ always increases with increasing $\theta$ up to 90°.

In order to understand the angle dependent magnetization data, we calculated the $\theta$ dependence of one representative spins $S_1$ (positive $S_z$ at $\mu_0 H$ = 0 T) under selected magnetic field up to 11 T. According to the symmetry argument and calculations, $S_{2z}(\theta)$ = $-S_{1z}(180°-\theta)$ and $S_{2y}(\theta) = S_{1y}(180°-\theta)$ so that results of $S_2$ spins are not shown here. All the field range can be classified into three groups according to $H//z$ configuration, $H<H_{c1}$, $H_{c1}<H<H_s$, and $H>H_s$, as shown in Figs. 5(a) and (d), (b) and (e), (c) and (f), respectively. For $\mu_0 H$ = 1 T curve, $S_1$ will slightly deviate from $z$-axis by tilting towards $y$-axis under rotating $H$. From $\mu_0 H$ = 2 and 3 T curves, $S_1$ will rotate counterclockwise first at lower angle, then rotate clockwise continuously, before 180°, it will rotate back to its original orientation. From $\mu_0 H$ = 4-7 T, the initial $S_1$ position is tilted away from $z$-axis towards negative $y$-axis. Then $S_1$ will rotate in clockwise continuously, pass the positive $y$-axis and point to negative $z$-axis in a canted way for $\theta$ = 180°. From $\mu_0 H$ = 8 to 10 T, $S_1$ will follow exactly the $H$ in lower and higher angle. In the intermittent $\theta$ values sandwiched by two $H = H_s$ transitions, it will deviate away from $H$ to the left until the condition $H = H_s$ is satisfied again. The intermittent angle range of unsaturation will become narrower with increasing $H$. At 11 T>$H_s$, the $S_1$ and $S_2$ will always follow the direction of $H$ that the $S_y$ and $S_z$ components display a sinusoidal behavior. In particular, for $\mu_0 H$ = 9 T, the two angles where $H = H_s$ are 48° and 132°, which are very close to the $H_s$ of 54° and 126° obtained from $\theta$ dependence of the $\rho_{zz}$ under 9 T at 2 K.

Finally, we put all the determined $H_{c1}$ value and $H_s$ values in experimental and theoretical simulations, together with the analytical Eq. (2) into a polar phase diagram, as shown in Fig. 6. All the data and theory are well consistent with each other, proving the validity of our local-moment Heisenberg model with proper material parameters in this system. With increasing $\theta$, the $H_{c1}$ will decrease to zero in a semi-eclipse. However, from our simulation, the first order nature of this transition quickly disappears in a very small angle, implying a critical end point for $\theta$ value smaller than 5° since we could not find hysteresis in *H* scan at this angle [20]. This point may deserve further study in experiments in the future.

## IV. CONCLUSION

In summary, high quality single crystals $MnBi_2Te_4$ are grown and its angle dependent magnetoresistance $\rho_{zz}$ (*H*) and $\rho_{zz}$ ($\theta$) are measured to derive the two reorientation transition values. Based on those critical field values, a local-moment Heisenberg model with proper exchange interaction and uniaxial magnetic anisotropy strength are developed. Our model Hamiltonian can well reproduce the experimental observations in the polar phase diagram through Monte Carlo simulation. The results on bulk $MnBi_2Te_4$ may readily be extended to a few layers by measuring the angle dependent magnetoresistance of layer sample and extracting the critical field values for the model Hamiltonian.

## ACKNOWLEDGMENTS

This work is supported by the Natural Science Foundation of China grant Nos. 52071041, 11974065, 11904348, 12004056. This work has been supported by Chongqing Research Program of Basic Research and Frontier Technology, China (Grant No. cstc2020jcyj-msxmX0263), Fundamental Research Funds for the Central Universities, China (2020CDJQY-A056, 2020CDJ-LHZZ-010, 2020CDJQY-Z006), Projects of President Foundation of Chongqing University, China (2019CDXZWL002). We would like to thank Miss G. W. Wang at Analytical and Testing Center of Chongqing University for her assistance.

**FIGURE CAPTIONS**

**FIG. 1.** MnBi$_2$Te$_4$ crystal and magnetic structure for (a). Single crystal X-ray diffraction pattern with (00*l*) indexed peaks of (b). Temperature dependence of the inverse magnetic susceptibility $\chi^{-1}$ in field-cooled mode is shown in (c). The dash lines represent the Curie–Weiss fitting. In-plane resistivity $\rho_{xx}$ (*T*) for (d) under $\mu_0 H$ = 0 T and 9 T for *H*//*z*. The inset of (b) shows an image of MnBi$_2$Te$_4$ single crystal peeled off with tape.

**FIG. 2.** Field dependence of magnetization for (a) out-of-plane and (b) in-plane *H* measured at selected temperatures. *H-T* phase diagram of MnBi$_2$Te$_4$ for (c) *H*//*z* and (d) *H*//*xy*.

**FIG. 3.** Schematic configurations for the magnetoresistance measurements under the (a) horizontally rotating *H* and (b) that under the vertically rotating *H*. Azimuthal angle $\varphi$ and polar angle $\theta$ are defined as the relative angles between *H* and *y* and between *H* and *z*, respectively. *H* dependence of the resistivity $\rho_{zz}$ at 2 K at selected (c) $\varphi$ and (d) $\theta$ values. $\theta$ dependence of the $\rho_{zz}$ (e) at 2 K at selected *H* (f) under 9 T at selected temperatures.

**FIG. 4.** Monte Carlo simulation results of *H* dependence of the spin components (a) $S_{1y}$, (b) $S_{2y}$, (c) $S_{1z}$ and (d) $S_{2z}$ from the two representative spins $S_1$ and $S_2$ with positive and negative $S_z$ at 0 T, respectively at selected $\theta$ values for *H* increase run.

**FIG. 5.** Monte Carlo simulation results of $\theta$ dependence of the spin components $S_{1y}$ and $S_{1z}$ at selected magnetic field of (a) and (d) 1-3 T, (b) and (e) 4-7 T, (c) and (f) 8-11 T, respectively in the angle increasing run.

**FIG. 6.** Polar phase diagram of crystal studied, which is determined by from Eq (2), and critical field values obtained from Figs. 2(a), 3(d), 3(e), 4(d) and 5(c).

**FIGURES**

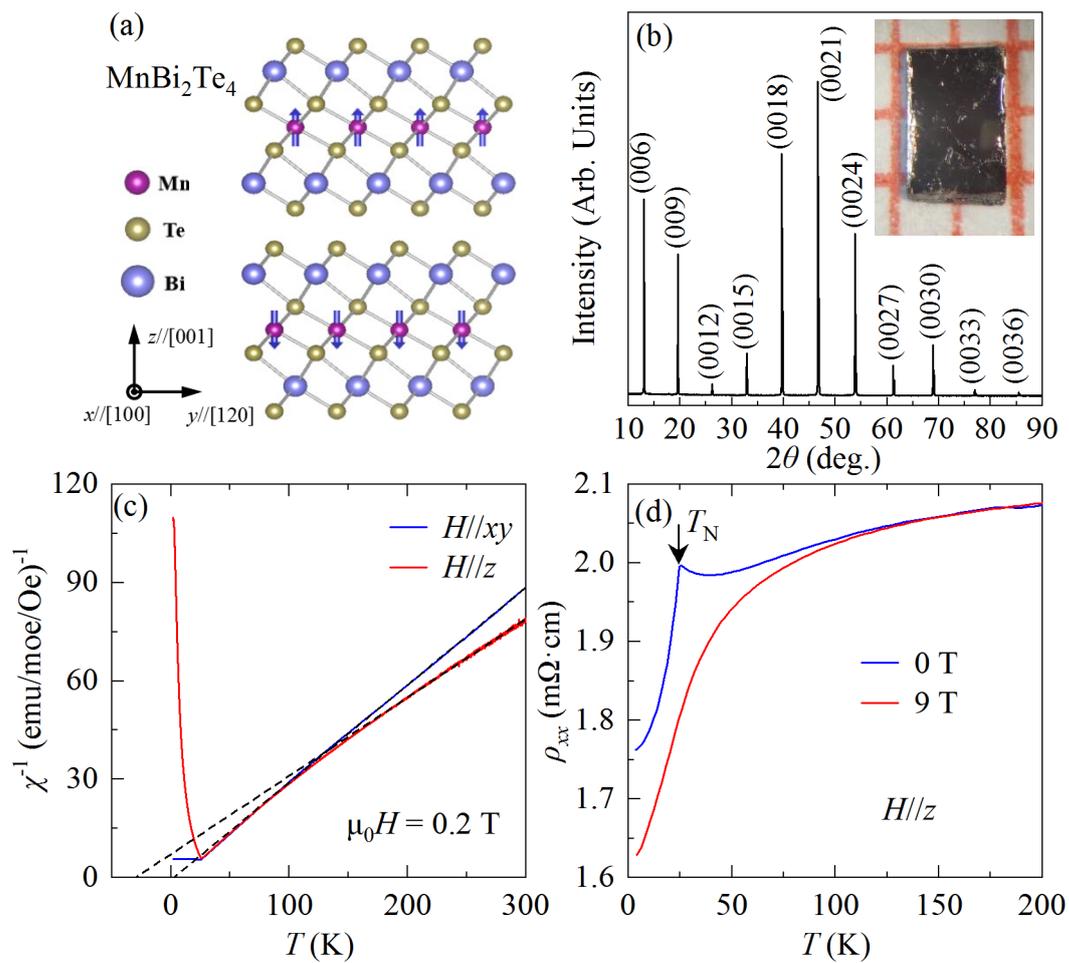

Figure 1 Ning Cao *et al*.

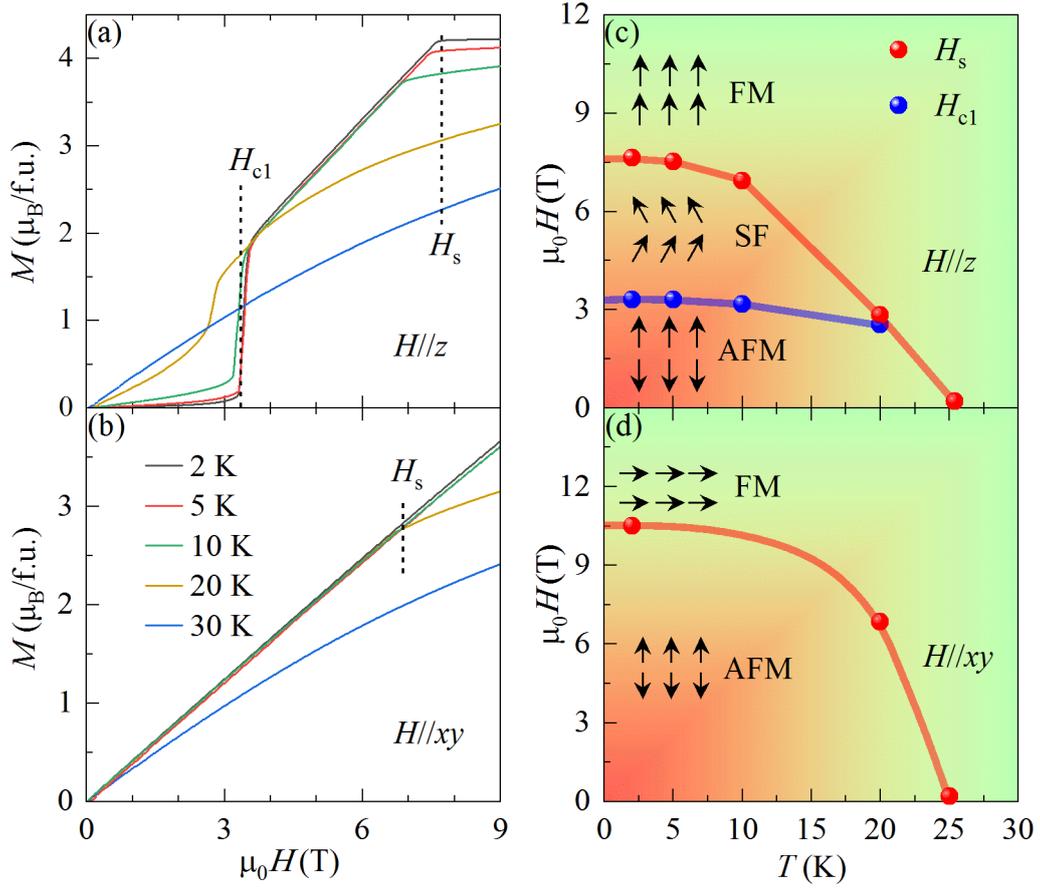

Figure 2 Ning Cao *et al*.

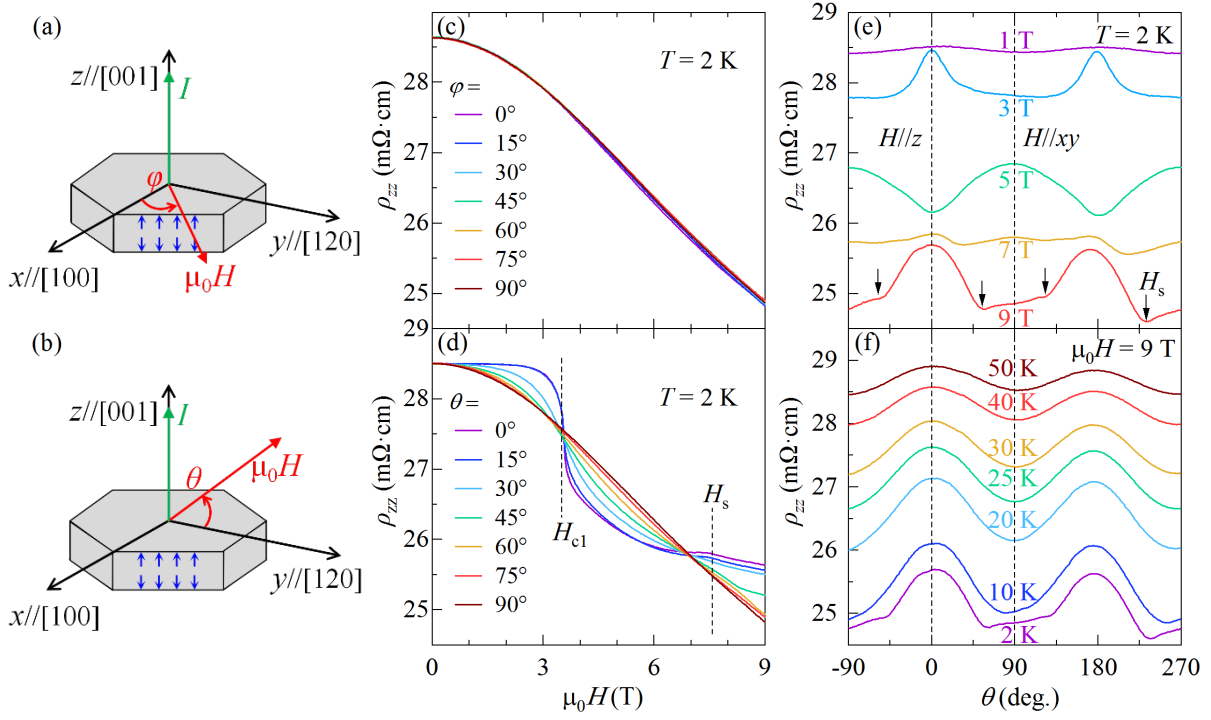

Figure 3 Ning Cao *et al*.

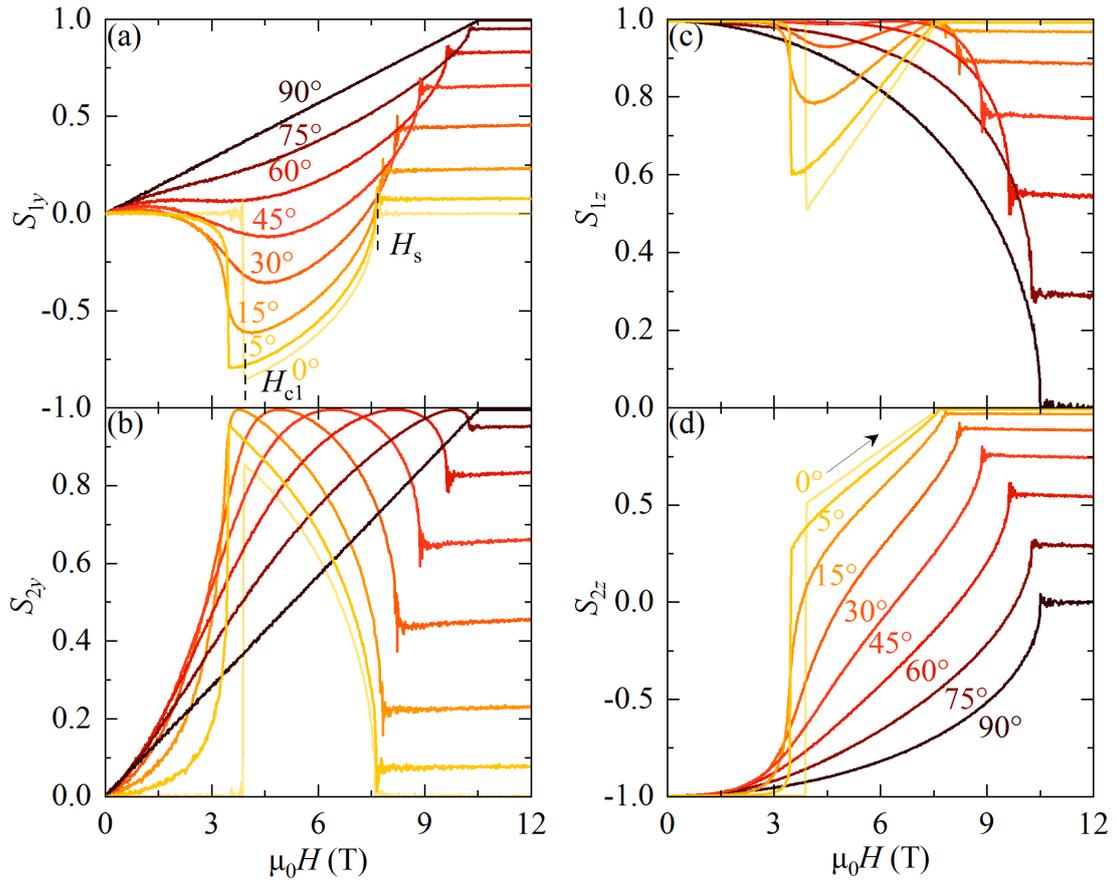

Figure 4 Ning Cao *et al*.

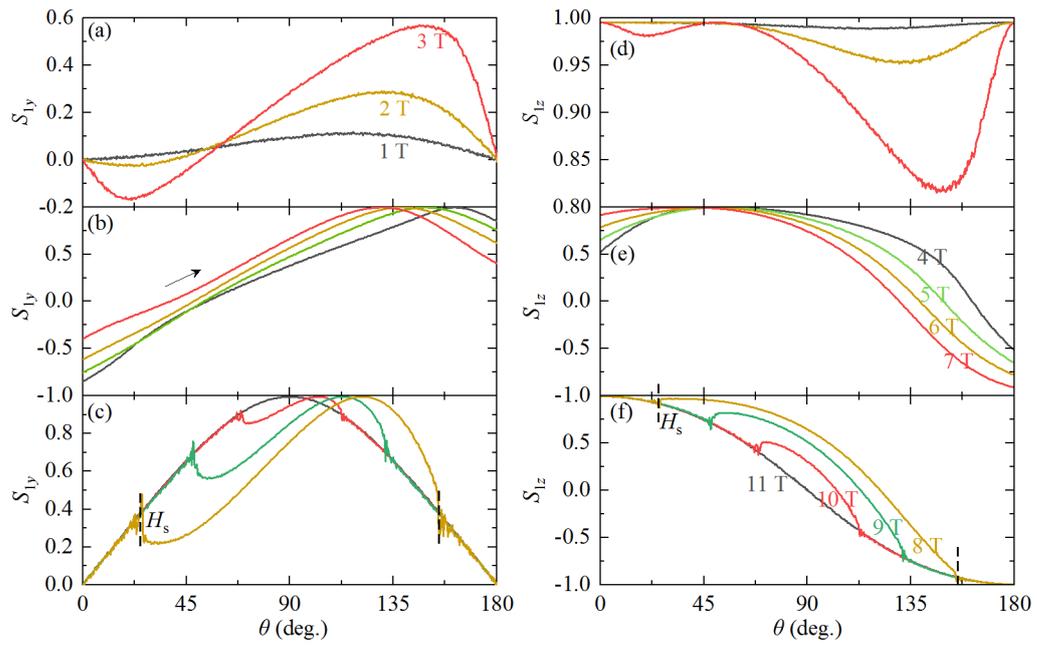

Figure 5 Ning Cao *et al*.

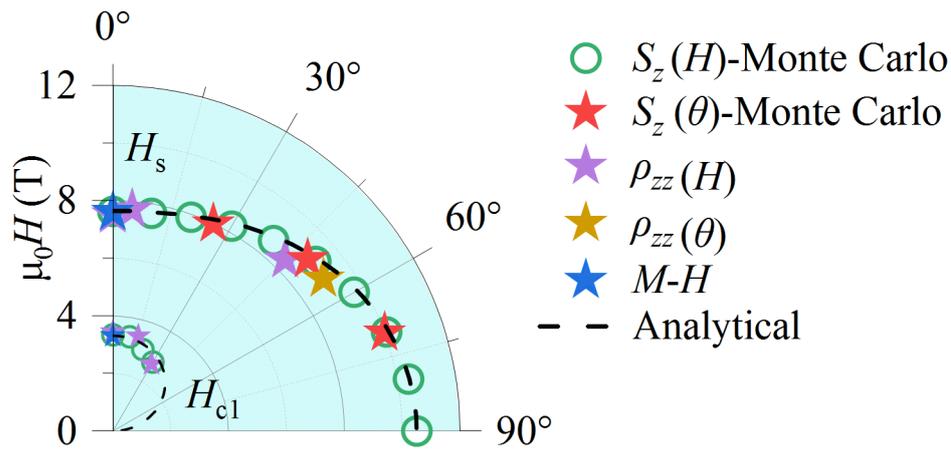

Figure 6 Ning Cao *et al*.